\documentclass[conference,10pt]{IEEEtran}
\usepackage[noadjust]{cite}

\usepackage{graphicx}
\usepackage{booktabs}
\usepackage{listings}
\usepackage{xurl}
\usepackage{amsmath,amssymb,amsfonts}
\usepackage{mathtools}
\usepackage{physics}
\usepackage{caption}
\usepackage{subcaption}
\usepackage{multirow}
\usepackage{xcolor}
\definecolor{uzlmain}{RGB}{0,75,90}
\definecolor{uzlpal6}{RGB}{198,220,226}
\usepackage{tcolorbox} 
\tcbset{colback=uzlpal6!50,colframe=uzlmain,boxrule=0.8pt,sharp corners=northwest} 

\usepackage{siunitx}
\usepackage[colorlinks=true,urlcolor=black,linkcolor=black,citecolor=black]{hyperref}
\def\BibTeX{{\rm B\kern-.05em{\sc i\kern-.025em b}\kern-.08em
    T\kern-.1667em\lower.7ex\hbox{E}\kern-.125emX}}

\usepackage[capitalise]{cleveref}
\usepackage{orcidlink}

\newcommand{\tool}{\textsc{Zebrafix}}
\newcommand{\cf}{Cipherfix}
\newcommand{\ch}{CipherH}
\newcommand{\cio}{cio}
\newcommand{\gofetch}{GoFetch}

\newcommand{\mbedtls}{\texttt{mbedTLS}}
\newcommand{\lsodium}{\texttt{libsodium}}

\newcommand{\eddsa}{\texttt{Ed25519}}
\newcommand{\chacha}{\texttt{ChaCha20}}
\newcommand{\sha}{\texttt{SHA512}}
\newcommand{\baseLXIV}{\texttt{Base64}}

\newcommand{\xsp}{XorShift128\raisebox{.25ex}{+}}

\newcommand{\uarch}{microarchitectural}
\newcommand{\uarchopt}{microarchitectural optimization}
\newcommand{\uarchopts}{microarchitectural optimizations}

\newcommand{\ctsc}{ciphertext side-channel}
\newcommand{\Ctsc}{Ciphertext side-channel}
\newcommand{\CtscTitle}{Ciphertext Side-Channel}
\newcommand{\ctscs}{ciphertext side-channels}
\newcommand{\Ctscs}{Ciphertext side-channels}
\newcommand{\CtscsTitle}{Ciphertext Side-Channels}

\newcommand{\sss}{silent store}
\newcommand{\SssTitle}{Silent Store}
\newcommand{\Sss}{Silent store}
\newcommand{\ssss}{silent stores}
\newcommand{\SsssTitle}{Silent Stores}
\newcommand{\Ssss}{Silent stores}

\newcommand{\dmps}{data memory-dependent prefetchers}
\newcommand{\Dmp}{Data memory-dependent prefetcher}
\newcommand{\Dmps}{Data memory-dependent prefetchers}
\newcommand{\DmpsTitle}{Data Memory-Dependent Prefetchers}
\newcommand{\dmping}{data memory-dependent prefetching}

\newcommand{\DMP}{DMP}
\newcommand{\DMPs}{DMPs}

\newcommand{\cswap}{\texttt{ct}-\texttt{swap}}
\newcommand{\Cswap}{\texttt{Ct}-\texttt{swap}}

\newcommand{\scname}{memory-centric}

\newcommand{\ScNameTitle}{Memory-Centric}

\renewcommand{\texttt}[1]{$\mathtt{#1}$}

\newcommand{\bheading}[1]{\vspace{5pt}\noindent{\textbf{#1.}}}

\newcommand{\appsection}[1]{%
  \section{#1}%
}
\crefname{appsec}{Appendix~Section}{Appendix~Sections}
\Crefname{appsec}{Appendix~Section}{Appendix~Sections}

\title{\tool{}: Mitigating \ScNameTitle{} Side-Channel Leakage via Interleaving}

\makeatletter
\def\@IEEEsectpunct{.\ \,}
\def\paragraph{\@startsection{paragraph}{4}{\z@}{1.5ex plus 1.5ex minus 0.5ex}%
{0ex}{\normalfont\normalsize\itshape}}
\makeatother

\begin{document}

\author{
  \IEEEauthorblockN{
    Anna Pätschke~\orcidlink{0000-0001-7828-2333},
    Jan Wichelmann~\orcidlink{0000-0002-5748-5462},
    Thomas Eisenbarth~\orcidlink{0000-0003-1116-6973} 
  }
    \IEEEauthorblockA{
    University of Luebeck, L\"ubeck, Germany\\
    \{a.paetschke,\,j.wichelmann,\,thomas.eisenbarth\}@uni-luebeck.de
  }
}

\maketitle

\begin{abstract}    

    Constant-time code has become the de-facto standard for secure cryptographic implementations. 
    However, some memory-based leakage classes such as \ctscs{} and \ssss{} remain unaddressed.
    Prior work proposed three different methods for \ctsc{} mitigation, for which one, the practicality of interleaving data with counter values, remains to be explored.
    To close this gap, we define design choices and requirements to leverage interleaving for a generic \ctsc{} mitigation.
    Based on these results, we implement \tool{}, a compiler-based tool to ensure freshness of memory stores.
    We evaluate \tool{} and find that interleaving can perform much better than other \ctsc{} mitigations, at the cost of a high practical complexity. 
    We further observe that \ctscs{} and \ssss{} belong to a broader attack category: \scname{} side-channels.
    Under this unified view, we show that interleaving-based \ctsc{} mitigations can be used to prevent \ssss{} as well.

\end{abstract}


\section{Introduction} 
\label{sec:intro}

With the discovery and exploitation of an ever-increasing number of side-channels, constant-time code has become a fundamental requirement for secure implementation of cryptographic algorithms.
Constant-time code does not contain any secret-dependent memory accesses or branches, and restricts the use of instructions whose timing behavior depends on their operands.
Techniques like constant-time selection allow to securely read a value depending on a sensitive condition, while always accessing the same memory addresses.
Hence, the memory access pattern does not exhibit any side-channel information that may be exploitable by an attacker.

Nevertheless, profound leakage evaluations by Barthe et al.~\cite{DBLP:conf/ccs/BartheBCCGGRSWY24} and by Vicarte et al.~\cite{DBLP:conf/isca/VicarteSNT0KF21} show that countering only timing leakage is not sufficient to protect the secrecy of cryptographic keys and data. 
Logical bugs in the configuration of trusted execution environments (TEEs) as well as current and upcoming microarchitectural optimizations negate the former security guarantees of constant-time code by adding new secret-dependent observable behavior. 
The most notable examples of \scname{} side-channels are \ctscs{}~\cite{DBLP:conf/uss/LiZWLC21, DBLP:conf/sp/LiWW0TZ22}, \ssss{}~\cite{DBLP:conf/asplos/FlandersSMGK24} and \dmping{} (\DMP{})~\cite{wang2025peekPreprint, DBLP:conf/uss/ChenWSFKPG24, DBLP:conf/sp/VicarteFPG0FK22}. 

\emph{\Ctscs{}}~\cite{DBLP:conf/uss/LiZWLC21, DBLP:conf/sp/LiWW0TZ22} exploit the deterministic memory encryption of contemporary TEEs.
To protect a workload from privileged system-level attackers, TEEs ensure that all data written to main memory is encrypted, and only decrypted within the processor during execution of the sensitive workload.
This encryption is typically hardware-based and comes with other security features like a write protection, that prevents an attacker from changing or moving ciphertext blocks, or otherwise interfering with the workload's execution.
Most available TEEs rely on a tweaked block cipher for their memory encryption, i.e., the ciphertext at a particular memory address depends on the plaintext, a secret key, and a tweak.
The tweak is commonly derived from the physical address.
This implies that writing a given plaintext to a given memory address always yields the same ciphertext.
AMD's \emph{Secure Encrypted Virtualization} (SEV) TEE~\cite{SEV-SNP_whitepaper} lacks read protection, so the attacker can freely read the ciphertext.
If the attacker now observes recurring ciphertext values at a particular address, they immediately learn that the same plaintext was written to that address.
This small information leak is sufficient to break constant-time implementations~\cite{DBLP:conf/sp/LiWW0TZ22}.

The authors of~\cite{DBLP:conf/sp/LiWW0TZ22} highlighted several software-level defenses, which all aim to add freshness to the plaintext:
First, by data address rotation that changes the address of a memory object on each write, second, by masking that adds a random mask to the plaintext before writing, or third, by interleaving counters and plaintext fragments to force a ciphertext change. 

The first two of these, data address rotation and masking, have already been tested, with mixed results; leaving the third one, interleaving, to still be explored.
The first defense, data address rotation, has been shown to work well for manual patches in well-defined structures like the kernel register state~\cite{DBLP:conf/sp/LiWW0TZ22}, but otherwise it was deemed impractical for automated application in general purpose programs~\cite{DBLP:conf/uss/WichelmannPW023}. 
The second defense, masking, was found to be more promising for general applicability:
In \emph{\cf{}}~\cite{DBLP:conf/uss/WichelmannPW023}, the authors used a combination of taint tracking and binary rewriting to identify secret-dependent memory writes and harden these writes by adding a mask.
A drawback of the presented masking implementation approach is the high overhead that is partly due to the use of binary rewriting, and non-local accesses to the mask buffers.
However, most of the slowdown is caused by the mask generation itself, which is inherent to any implementation of masking and cannot be avoided.

Since the results of the first two defenses (data address rotation and masking) are not entirely convincing, the third one remains to be investigated, which is interleaving of secret data with counters. 
The idea of interleaving is eminently promising:
When dividing a typical 16-byte cipher block into two halves, the resulting 8-byte counter half guarantees $2^{64}$ unique ciphertexts before a repetition can occur, without requiring costly random mask generation.
Interleaving has been tested on a small scale for oblivious RAM~\cite{DBLP:conf/sp/WichelmannRPE24}, but the results do not indicate the general suitability of interleaving for \ctsc{} protection. 

In the first part of this work, we close the aforementioned gap:
We implement \tool{}, a proof-of-concept compiler plugin which automatically hardens a program against \ctsc{} leakage via interleaving.
For this, we identify the hurdles that need to be overcome, review relevant design decisions when implementing a \ctsc{} mitigation, and outline necessary preconditions.
We evaluate \tool{} on a set of cryptographic primitives, and show that results go both ways:
We observe that interleaving indeed performs much better than masking, while achieving a higher level of security. 
On the downside, we highlight several critical practical issues of automated interleaving, which are difficult to work around for contemporary compilers and non-type-safe programming languages.

In the second part of this work, we show that \ctscs{} are actually part of a broader category of \emph{\scname{} side-channel attacks}: 
Data that has been written to memory once leaks secret information through subsequent accesses to the unprotected data.
By generalizing the leakages to \scname{} instead of fixing each leakage in isolation, future mitigations can benefit from existing knowledge and proven effective strategies for developing a mitigation.
One illustrative example from the category of \scname{} side-channels is the upcoming \uarchopt{} \emph{\ssss{}}~\cite{DBLP:conf/isca/LepakL00}, which discards stores that would not result in a change to the memory contents.
Consequently, this breaks a vital assumption of constant-time code, namely that the behavior of stores is independent of the value written.
Another related example is \emph{\dmping{}} (\DMP{})~\cite{DBLP:conf/isca/VicarteSNT0KF21, intelDdp}, where the CPU prefetches data depending on (potential) pointers detected in recently accessed memory. 

We discuss similarities and differences between \ctscs{} and \ssss{}, and analyze the suitability of existing analysis and mitigation tools.
For example, a performant compiler-level mitigation against \sss{} leakage is presented in \emph{\cio{}}~\cite{DBLP:conf/asplos/FlandersSMGK24}.
The addition of a memory write with a provably distinct value between two consecutive memory writes effectively prevents silencing a store.
The \cio{} approach does not cover \ctsc{} leakage.

On the side of \ctsc{} analysis tools, \emph{\ch{}}~\cite{DBLP:conf/uss/DengLTWYZ23} aims to find leaky memory writes by reasoning whether two consecutive memory writes of the same secret variable to the same address can hold different values.
We observe that this approach has a limited analysis scope in terms of \ctsc{} leakage and would result in false negative results for code that is protected with a \cio{}-style \sss{} mitigation approach. 
There are currently no dedicated \DMP{} leakage mitigations; yet, a \cf{}-style masking approach is applicable to protect secret data at the cost of introducing high overheads.
In light of this, we point out how existing \ctsc{} defenses can be generalized to mitigate \sss{} leakage.
However, adapting interleaving to mitigate the entire class of attacks, including \DMP{}, is non-trivial and invites future research focusing on more holistic mitigations.

\subsection{Contribution} 
In summary, we make the following contributions: 
\begin{itemize}
    \item We review design decisions for \ctsc{} defenses and develop \tool{}, a proof-of-concept mitigation tool based on interleaving.    
    \item We evaluate the effectiveness of interleaving regarding performance overhead and security. 
    \item We provide recommendations for implementing generic defenses against \scname{} side-channel leakages in constant-time code.
    \item We discuss strategies for augmenting interleaving as a mitigation against other \scname{} side-channel leakages.
\end{itemize}

Our source code is available on GitHub at the following URL: \url{https://github.com/UzL-ITS/zebrafix}. 

\bheading{Outline} 
We provide background on the different leakage channels in~\Cref{sec:background} and establish interleaving as a \ctsc{} mitigation in~\Cref{sec:prev-leak}.
We continue with a comparison of different mitigation implementation layers in~\Cref{sec:leakdeflevels}.
Then, in~\Cref{sec:impl}, we present our \tool{} implementation, which we evaluate in~\Cref{sec:eval}.
We explain the interconnections between \ctsc{} leakages and \sss{} leakages in~\Cref{sec:prev-sss}.
We discuss potential enhancements and general technical limitations in~\Cref{subsec:limitations} and set our work in context with related work in~\Cref{sec:related-work}. 
We conclude with~\Cref{sec:conclusion}.
An additional discussion of \dmping{} in relation to interleaving can be found in~\Cref{app:dmp}.

\section{Leakage Preliminaries}
\label{sec:background}

Through logical flaws, microarchitectural optimizations or other channels, the secrecy of data that is written to memory without further protection is endangered.
While the term \scname{} initially included data-at-rest leakage that is not instruction-centric in~\cite{DBLP:conf/isca/VicarteSNT0KF21}, we generalize the term to also include instruction-centric leakages that originate in unprotected data-at-rest.
Information that may be disclosed in such \scname{} side-channel attacks includes whether a computed value is equal to another value that has been previously stored in the architectural state of main memory. 
A prime example for a constant-time primitive that was presumed to be secure, yet is susceptible to \scname{} leakage, is the constant-time swap.
We first introduce the properties of the constant-time swap and then present various \scname{} leakage scenarios on that example.
 
\subsection{Constant-Time Swap}
\begin{figure}[t]
    \centering
    \begin{subfigure}[b]{0.45\textwidth}
        \centering
        \includegraphics[width=.7\textwidth]{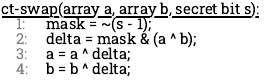}
        \caption{Simplified procedure of a constant-time swap (\cswap{}). Depending on the value of a secret decision bit \texttt{s}, the values in \texttt{a} and \texttt{b} are swapped (\texttt{s = 1}), or left as-is (\texttt{s = 0}).}
        \label{fig:cswap:code}
    \end{subfigure}
    \par\bigskip
    \begingroup
    \setlength{\tabcolsep}{12pt}
    \begin{subfigure}[b]{0.45\textwidth}
        \centering
        \includegraphics[width=\textwidth]{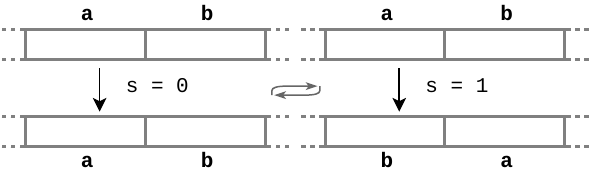}
        \caption{Simplified illustration of \cswap{} based on~\cite{cipherfixSlides}. The memory contents \texttt{a} and \texttt{b} at their respective two memory addresses (depicted by boxes) are not swapped if the secret bit was not set (\texttt{s = 0}, left case). If the secret bit was set (\texttt{s = 1}, right case), the contents in memory are swapped after the write operation. The curved arrows represent a potential swap of the \cswap{} procedure (line 3 and 4 in~\labelcref{fig:cswap:code}).}
        \label{fig:cswap:pic}
    \end{subfigure}
    \endgroup
    \caption{Constant-time swap of two arrays \texttt{a} and \texttt{b}.}
    \label{fig:cswap}
\end{figure}

Constant-time code has been established as de facto standard for cryptographic implementations.
A prominent example is a constant-time swap (\cswap{}): Depending on a secret bit \texttt{s}, the contents of two arrays \texttt{a} and \texttt{b} are swapped.
The procedure is shown in~\Cref{fig:cswap:code}.
The secret bit is used for building a mask value, extended to a machine-sized word that either resembles \texttt{0x00...00} or \texttt{0xFF...FF} (line 1).
With the mask, an intermediate value \texttt{delta} is set to propagate the information whether \texttt{a} and \texttt{b} are to be swapped (line 2).
Thereby, in lines 3 and 4, the variables \texttt{a} and \texttt{b} are always updated, independent of whether a swap occurs. The write accesses thus always go to addresses that do not depend on the underlying data.
A graphical representation of this is given in~\Cref{fig:cswap:pic}.
\Cswap{} does not contain secret-dependent memory accesses or branches.
However, there are many ways to still leak secrets from this classical constant-time example, as presented in the following subsections.
While in some scenarios secret data can be leaked without the influence of attacker-controllable data, some leakage classes require suitably crafted calculations and preparations in order to be able to attack secret information from constant-time implementations.

\subsection{\CtscsTitle{}}
\label{subsec:ctscs}

\Ctsc{} leakage occurs in VM-TEEs with deterministic memory encryption, where an attacker with root access to the machine is able to observe ciphertexts of the data written to main memory by a victim VM~\cite{DBLP:conf/sp/LiWW0TZ22, DBLP:conf/uss/LiZWLC21}.
An example of this are VMs protected with AMD SEV-SNP~\cite{SEV-SNP_whitepaper}, where the hardware-assisted memory encryption is based on a secret key, physical address-dependent tweaks, and the plaintext to be written to an address.
For each encrypted 16-byte block, there is no additional freshness added when the same data is rewritten to the same address.

\begin{figure}
    \centering
    \includegraphics[width=\linewidth]{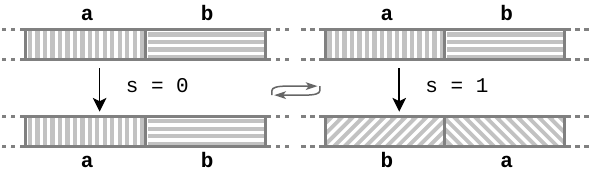}
    \caption{\Ctsc{} leakage in \cswap{} based on~\cite[p. 3]{cipherfixSlides}. Ciphertexts are depicted in different shades of the memory address block. Although an attacker can only see ciphertexts of the values in arrays \texttt{a} and \texttt{b}, a ciphertext \emph{change} leaks the value of the secret decision bit \texttt{s} during the \cswap{}.}
    \label{fig:cswap-ctsc}
\end{figure}

The authors of~\cite{DBLP:conf/sp/LiWW0TZ22} present two attacks that are based on the missing freshness of ciphertexts, dubbed \emph{dictionary attack} and \emph{collision attack}, whereby both rely on monitoring repeated write accesses to the same memory address.
The dictionary attack enables the attacker to map a set of ciphertexts to a set of known plaintexts to learn information about secret data.
In the stronger attack scenario, the collision attack, the attacker knows that there are only two different ciphertexts at a given address.
From this, the attacker can infer whether the data being written is the same as the data already present at that address by learning an equality predicate of the data.
When the attacker monitors the respective architectural state after multiple memory writes to certain addresses, they can infer whether the plaintext changed or not after each write access: The ciphertext at that address changes \emph{if and only if} the plaintext is different. 
A graphical representation of \ctsc{} leakage through collision attacks is given in~\Cref{fig:cswap-ctsc}.
\Ctsc{} mitigations are further investigated in~\Cref{sec:prev-leak}.

\subsection{\SsssTitle{}}
\begin{figure}
    \centering
    \includegraphics[width=\linewidth]{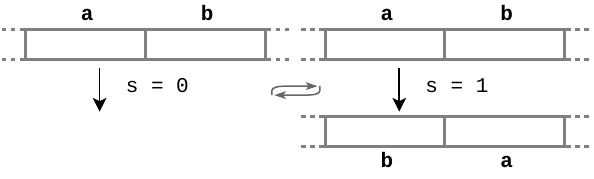}
    \caption{\Sss{} leakage in \cswap{}. If the secret decision bit \texttt{s} is unset, the value to be written to the memory address is the same as the already contained one. Therefore, \sss{} optimizations drop the memory write. The constant-time properties of the \cswap{} are violated as an attacker can observe whether a memory write happened or not.}
    \label{fig:cswap-sss}
\end{figure}

\Ssss{} are a microarchitectural optimization aiming to reduce the number of stores to be executed~\cite{DBLP:conf/isca/LepakL00}.
If a store to a specific address is about to write the value already present at that address, the store is aborted.
An illustration of this potential leakage is given in~\Cref{fig:cswap-sss}.
Apart from being documented for RISC-V~\cite{riscvSilentStores}, \ssss{} have also been observed on x86 for suppressing writes of certain values~\cite{intelSilentStores}.

An attacker has to be able to observe the \uarch{} state of system components (e.g., cache state, execution time or store queue pressure) to derive whether a store has been silenced or not.
\Ssss{} can either lead to similar leakage as the \ctsc{} by observing whether data at a certain address changes or not, or it can lead to leakage by observing whether an attacker-controlled value and a secret value are equal and lead to the suppression of a \sss{}.

The leakage from this \uarch{} optimization can be circumvented by ensuring that no memory write to a specific address contains a value that is already present at that address, e.g., by adding a write with new data in between or applying software-based probabilistic encryption to the data before writing it to memory~\cite{DBLP:conf/asplos/FlandersSMGK24}.

\section{Preventing \CtscTitle{}s}
\label{sec:prev-leak}

To counter \ctsc{} leakage, masking has been thoroughly explored as a binary-level mitigation, while interleaving has received little attention yet.
In the following, we describe interleaving as a ciphertext side-channel mitigation in depth and discuss associated design decisions.

\subsection{Threat Model}
\label{subsec:threatmodel}

We assume the standard \ctsc{} leakage scenario, where a virtual machine (VM) is running on external shared hardware (e.g., a cloud machine).
The hypervisor is considered malicious and tries to extract secret data from the VM.
To avert such attacks, the VM is protected with memory encryption, that uses a tweaked block cipher with physical address-based tweaks.
We further assume that the VM state save area and kernel data structures are protected, leaving user space applications vulnerable.
The attacker has full read access to the encrypted memory and can monitor the microarchitectural state of shared components like CPU caches.
An example for this scenario is AMD SEV-SNP. 

Applications running inside the VM are assumed to be constant-time (or constant-time after disabling certain microarchitectural optimizations), i.e., the code does not contain any secret-dependent memory accesses or branches that would be exploitable via a cache attack.
Constant-time in the TEE scenario also excludes intra-cache line leakages~\cite{DBLP:journals/ijpp/MoghimiWES19, DBLP:conf/ccs/SieckBW021, DBLP:journals/tches/SieckZBCEY24}.
We consider physical or power-related attacks as well as transient execution attacks out-of-scope.
Furthermore, we assume that the constant-time code does not contain other vulnerabilities like buffer overflows, and that the compiler preserves the constant-time properties.

\subsection{Interleaving as Leakage Defense}
\label{subsec:leak-defs} 

While there are plans to add hardware-based read protections in AMD SEV-SNP for upcoming CPU generations~\cite{AMD-ciphertext-hiding}, the currently available processors need a software approach for protection.
There are three software techniques that can be employed to prevent \ctsc{} leakage~\cite{DBLP:conf/sp/LiWW0TZ22, DBLP:conf/uss/WichelmannPW023}, each taking a different route at circumventing ciphertext determinism:
First, to prevent repeating plaintexts at the same address, the reuse of memory locations can be limited.
Second, plaintexts can be masked with random values to introduce software-based nondeterminism.
Masking adds a random bit string before stores, which needs to be subtracted on loads.
This raises certain issues, like the quality of the random number generator, which strongly correlates with the performance overhead. 
Both, rotating addresses of data and masking, were found to incur considerable performance overhead in practical implementations~\cite{DBLP:conf/sp/LiWW0TZ22, DBLP:conf/uss/WichelmannPW023}.

The third approach is similar to the masking approach, as it enforces constantly changing plaintexts, but it does so by interleaving the data with a counter:
Each encryption block is divided into two halves, where the first one is a counter that is incremented on each memory store. 
As \ctsc{} leakage arises from repetition of data that is written to memory, an effective mitigation is to add freshness to each written value.
While masking needs to ensure that plaintext and mask do not cancel each other out and thereby lead to repeating ciphertexts, strong randomness is not necessary in the general case to prevent plaintexts from repeating, since the memory encryption provides sufficient diffusion even if only a single plaintext bit changes at a time.
Thus, while masking needs a certain level of good randomness due the unpredictability of the protected values, an approach that is based on interleaving can omit the per-write randomness generation.
Reserving a certain part of each plaintext block for a deterministic counter that is only initially seeded by strong randomness is just as secure, and promises better performance.

\paragraph{Protection Scope} 
In line with previous work, protected data types include primitive integers, and arrays or structs containing those types.
As noted in~\cite{DBLP:conf/uss/WichelmannPW023}, pointers do not need protection:
The memory location of variables can be known, and secret-dependent memory accesses are eliminated by the constant-time code.
Typically, the protected data resides on the stack, the heap or in global variables.

\paragraph{Block and Counter Size}
Common memory encryption schemes use AES, which has a 128-bit block size.
Thus, each plaintext block has 16 bytes.
The amount of bytes reserved for the counter is a security parameter, as it determines the minimum number of memory writes for which no collisions are possible.
Given that the native integer size of most current systems is 64-bit, it is reasonable to divide each 16-byte block evenly into two 8-byte chunks.
This guarantees $2^{64}$ writes of a 16-byte block without collisions.

\begin{figure}
    \centering
    \includegraphics[width=\linewidth]{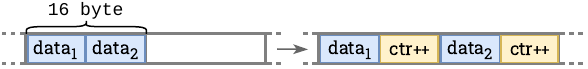}
    \caption{Interleaving data and counter values in memory. In order to ensure freshness in each 16-byte block in memory, the blocks are split into 8-byte counter and 8 byte available for data. On each memory write, the counter value is incremented.}
    \label{fig:zebra}
\end{figure}

Increasing the counter size yields an even larger security margin, but would lead to splitting 8-byte numbers, which would increase implementation complexity and overhead considerably.
This issue also applies when the counter size is decreased, as the opened space is smaller than 8 bytes.
We thus conclude that a counter size of 8 bytes is optimal in terms of security and implementation complexity.
An example for interleaving with 8-byte counters is shown in~\Cref{fig:zebra}.

\subsection{Realization Considerations} 
A realization of interleaving needs to ensure that counters and data are written \emph{simultaneously}; else, adjacent writes of the same value would be exposed.
Such 16-byte writes can typically be achieved through vector extensions.
Additionally, the extensive changes of data layout require instrumentation of all loads and stores to the affected addresses, as well as allocations and address computations. 

Specifying parts of the code to be protected can help improve the performance and efficiency of automated \ctsc{} defenses.
A mitigation may generally pursue two approaches for indicating what should be protected:
First, by marking certain variables as \emph{secret}, or second, by annotating code sections or functions.

\cf{} takes the first approach:
The authors modify the application source code to call a special function, passing each secret as parameters.
They then employ dynamic taint tracking to find all instructions accessing that variable.
However, using dynamic methods for secrecy tracking is susceptible to insufficient coverage.
Missing a code path during analysis eventually leads to bugs in the protected application, where an instruction accesses a protected memory location under a wrong assumption.
Static taint analysis tools may avoid this issue. 
Yet, static analyses often overapproximate results and can still be incomplete in the tracking of exact data types, leading to error-proneness in the instrumentation. 
Any secrecy tracking that leads to hybrid (instrumented and non-instrumented) usage of global variables endangers functional correctness of the program, as detailed in~\Cref{subsec:limitations}.

To avoid this hybrid usage, we take the second approach in \tool{}, i.e., comprehensive function annotation.
We protect all variables and computations in the call tree underlying the outmost caller function of the cryptographic routines, reducing the taint tracking problem to a simpler and broader call graph analysis that leads to protection of all called functions and all allocated heap memory.
Though instrumenting all functions in the call graph clearly overapproximates the protection scope (comparable to heavy overtainting), it simplifies the implementation and greatly reduces the error-proneness. 
While the intent of precise information flow tracking is reduction of the runtime performance overhead, we found that \tool{} performs quite well, despite the overapproximation. 

\section{Defense Implementation Layers} 
\label{sec:leakdeflevels}

As previous research indicated, \ctsc{} defenses can be realized at different implementation layers:
The ad-hoc patches to WolfSSL~\cite{wolfssl21patches} and the Linux kernel~\cite{DBLP:conf/sp/LiWW0TZ22} were made to the source code, while \cf{}~\cite{DBLP:conf/uss/WichelmannPW023} directly modified the compiled binaries through binary rewriting.
In the following, we provide guidance on a suitable implementation layer for an automated mitigation.
We accumulate knowledge gained from existing side-channel defenses, explore possible solutions and discuss practical limitations.
For compiler-based approaches, we generally use terminology from the LLVM project~\cite{DBLP:conf/cgo/LattnerA04}, which has long become a standard tool for side-channel defense implementations.
However, our conclusions also apply to similar compiler infrastructures.

\subsection{Binary Rewriting}
In binary rewriting, the modifications are directly applied on an existing binary.
This approach has the advantage of being able to harden an application specifically for deployment on a vulnerable system, e.g., an SEV-SNP protected VM, without needing recompilation.
Another advantage is the ability to protect not only a single application binary, but its dependencies as well, which may be too complex to compile and deploy separately (e.g., libc).
The most significant challenges of binary rewriting are the stability of the modified binary, and runtime performance~\cite{DBLP:conf/uss/WichelmannPW023}.
While binary rewriting is generally possible for typical instrumentation tasks, \ctsc{} defenses require very extensive changes, pushing any binary rewriting framework to its limits~\cite{DBLP:conf/sp/DineshBXP20, DBLP:conf/uss/BartolomeoMP23}.

In general, binary rewriting is highly complex due to the lack of information and control about register usage and stack layout.
To protect local variables against \ctsc{} leakage, a tool needs to correctly identify their location and size, which is complicated by runtime stack realignment and stack slot reuse. 
Other compiler optimizations like jump tables restrict a rewriting framework's ability to modify machine code in place; instead, most frameworks resort to trampoline approaches where execution is redirected to a newly added code segment~\cite{andriesse2018practical}.
Performance-wise, binary rewriting comes with penalties for frequent jumps between code segments, which have bad locality.
Another important issue is the inherent lack of available registers, as these are typically already in use by the original code.
When an instruction is instrumented, some registers have to be freed, which usually involves another store/load cycle and corresponding \ctsc{} protection.

Finally, precise variable tracking is mandatory, as pointer and type information is not available. 
Missing a particular code path may lead to underapproximation, which in turn results in spurious runtime crashes due to unexpected data layout.
In summary, while binary rewriting works in principle, it was deemed too unstable for production use~\cite{DBLP:conf/uss/WichelmannPW023}.

\subsection{Compiler: Machine IR/Back End}
Instead of using binary rewriting, \ctsc{} mitigations can be implemented as a compilation pass.
The closest layer to binary rewriting is the back end of the compiler.
In LLVM, the code generator works with an architecture-specific intermediate representation (IR), called \emph{machine IR} (MIR).
MIR looks very similar to the final binary, but still has symbolic offsets, which greatly simplifies insertion of new code. Additionally, some information from higher compiler layers is available, e.g., the stack layout and limited pointer or type information. 

In the back end, there are different stages of lowering the architecture-independent IR to MIR.
In order to work with the back end code that already contains all the memory writes present in the compiled binary, the mitigation passes must operate after the register allocation is completed.
Therefore, the instrumentation must either perform the same steps to free registers as the binary rewriting, or use an option to ignore certain registers at the cost of a global performance impact.
Instrumentation must also take care of preserving architectural state like the status flags register, which can be changed by arithmetic operations such as mask application or counter increments. 
Similar to binary rewriting, tracking of variables may be necessary, if the MIR does not offer detailed pointer and type information from earlier compilation steps. 
We conclude that MIR is generally the better choice over binary rewriting in terms of stability and performance, if there are no dependencies that should be instrumented as well.

\subsection{Compiler: LLVM IR/Middle End}
In the middle end, all source code has been translated into an architecture-independent IR.
In LLVM IR, all semantic information from the original program is available, and introducing new variables and modifying instructions can be done easily and without side effects.
Machine-specific optimizations did not yet run, so the compiler is likely to produce efficient machine code for the \ctsc{} defenses.
Architecture-specific operations like vectorized 16-byte memory writes are available as intrinsics, which are later translated into their machine-layer counterparts. 
The correct translation to architecture-specific instructions with protected 16-byte writes can then be verified with binary-level analysis.

Restrictions come from handwritten and inline assembly, which is only available as an opaque block in the IR passes, and cannot be easily adjusted there without lifting it into IR.
As a workaround, a hybrid approach may harden these remaining code snippets in an MIR pass.
Another issue are stack spills (also referred to as register spills) during lowering the IR into MIR: When the register allocator finds that it does not have sufficient registers, it may choose to temporarily store values on the stack.
These stack spills only become visible at the MIR layer, where they would need special treatment. 
Special treatment can include writing the data to vector registers instead in order to avoid \ctsc{} leakage.

\bheading{Summary} 
In order to benefit from eased adjustments to memory layouts, instrumentation needs to be on middle end level.
The advantages of back end level instrumentation only become apparent if the instrumentation happens after the register allocation to be able to protect all memory writes.
Yet, at the back end stage, it is already sufficiently complicated to adjust the memory layout for interleaving, making it comparable to binary instrumentation.
We conclude that hardening \ctscs{} at the IR layer promises the best final performance; and combined with a binary-level assessment, an appropriate security assurance is provided.

\section{\tool{} Implementation}
\label{sec:impl}

\begin{figure}
    \centering
    \includegraphics[width=0.85\linewidth]{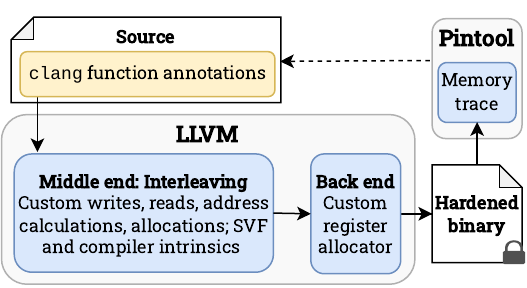}
    \caption{\tool{} toolchain. The developer marks sensitive workloads via function annotations. Then, the middle end passes add the main interleaving capabilities to the code. In the back end, the register allocation is adapted to prefer spilling to vector registers over spilling to the stack. With memory traces, the binary can be checked for leaky writes.}
    \label{fig:zebra-framework}
\end{figure}

Motivated by the considerations in the previous sections, we outline the conception of a compiler-based approach at LLVM middle and back end level to interleave data with counter values for adding freshness to each memory write.
In contrast to other side-channel mitigations that are implemented on MIR level only, we also include the LLVM IR level for supporting more optimizations and generalization between different architectures as long as the resulting instrumented binary is still safeguarded.
While this leads to reliance on certain properties of the compilation process, mitigations can benefit performance.

As a case study for evaluating the described interleaving approach, we implement a proof-of-concept, dubbed~\tool{}.
With that, we showcase the versatility of interleaving as a side-channel defense.
\tool{} is implemented as a set of LLVM~16\footnote{LLVM~16 is at the time of writing the newest LLVM version supported by SVF.} out-of-tree passes with SVF~\cite{DBLP:conf/cc/SuiX16} pointer analysis, with \num{2600} LOC in total. 
An overview of the \tool{} toolchain can be found in~\Cref{fig:zebra-framework}.
The developer who wants to protect a certain function against \ctsc{} leakage marks it with a \texttt{clang} function attribute.
The attribute is automatically propagated through the call tree, leading to instrumentation of all called child functions as well.
Indirect calls and unsupported components like inline assembly are printed for further inspection by the developer during the call tree instrumentation which is based on the LLVM call graph.
We merge all bitcode files that are generated for whole-program analysis and then instrument the merged bitcode file.
A discussion about the interaction of instrumented and non-instrumented (e.g., dynamically linked) code can be found in~\Cref{subsec:limitations}.
\tool{} supports the block size 16 byte with data chunks up to size $16/2=8$ byte.
An example of data chunks smaller than 8 byte can be found in~\Cref{fig:zebra-dummy}.

\subsection{Compiler-Level Transformations} 
\paragraph{LLVM Middle End}
We base our mitigation implementation mainly on LLVM IR passes, after all IR optimization has run.
This allows us to work on optimized LLVM IR, but before register allocation.
To keep the approach modular and to ease combination with other mitigations, we adjust the data layout via IR passes, instead of manipulating allocators and memory accesses via a runtime library.

Implementing the mitigation on LLVM IR level succeeds with the current standard configuration of LLVM to produce interleaved memory writes when the compilation process is otherwise adjusted correctly: 
For a working transformation with \tool{}, vector registers need to be available on the target, and compile steps prior to the instrumentation passes must avoid using them.
As the interleaved counter and the data must be written simultaneously per 16-byte block, vector writes are inserted.
For each primitive type, we provide a crafted struct that contains the suitable data type, some dummy elements and the 8-byte counter.
Adding own primitive types that have a suitable length would break compatibility with existing LLVM versions, including the analysis and optimization infrastructure\footnote{\url{https://llvm.org/docs/ExtendingLLVM.html\#adding-a-new-type}}.
Working with existing primitive types in structs combined with inserting vector writes allows us to circumvent the insertion of own types so that we can also take advantage of the highly optimized LLVM infrastructure. 
Moreover, with the flexibility of using structs to fill data into memory, we can comprehensibly adjust to other leakage scenarios stemming from other current and upcoming \uarchopts{} (see~\Cref{sec:prev-sss} and~\Cref{app:dmp}).

\paragraph{LLVM Back End}
In addition to the LLVM IR passes, we adjust the register allocator for x86-64 targets to decrease the number of stack spills of potentially secret data during function calls or because of high register pressure inside of functions.
For that, we integrate a register allocator enhancement that spills to vector registers instead of the stack, proposed by Matthias Braun\footnote{We adapted \url{https://discourse.llvm.org/t/rfc-spill2reg-selectively-replace-spills-to-stack-with-spills-to-vector-registers/59630/15}.}.
If there are not enough vector registers available for saving values, we print a warning so that the developer can check whether the write actually contains secret-dependent data or just irrelevant data like a pointer that is written to the stack.
The check is further enhanced with a binary-level assessment tool.
For the sake of maintainability and extensibility of \tool{}, we chose not to implement interleaved stack spilling, which would have required deep changes to the stack allocation.

\subsection{Adjusting Compiler Intrinsics}   
Another challenge for an instrumentation tool are compiler intrinsics.
These look similar to regular function calls, but are internal to the compiler and replaced by optimized machine code during code generation.
Intrinsics may be called by the developer, but often are inserted by the compiler in place of standard library functions or typical programming idioms.
For the correct handling of compiler intrinsics that alter memory contents, the necessary changes to the memory layout need to be clearly defined.
This especially includes knowledge of the types (and also the primitive types contained in aggregate types like arrays and structs) of data and allocations.

Examples of memory-altering intrinsics are \texttt{memcpy} and \texttt{memset}, which we replace with own implementations in \tool{}.
As the 16-byte blocks that are built with \tool{} contain varying amounts of data (e.g., 1 byte for chars, 4 byte for short integers, 8 byte for long integers), it does not suffice to adjust the amount of bytes copied or set by a fixed factor.
Instead, the correct number of elements of a specific primitive type to be copied has to be determined.
In the \tool{} implementation, we use SVF to infer the types of (source and) destination objects for the memory-altering intrinsics. 
With that, the data can in most cases properly be extracted such that a suitable interleaved struct for the copy or set destination can be generated.
The struct is then used for inserting new instrumented writes to memory.
In addition, the counter values must not only be copied, but also incremented to prevent leakage from multiple write accesses with the same counter value.
Thereby, even with multiple copies of the buffer, the counters are incremented to avoid repeating ciphertexts.

\subsection{Instrumentation Aspects for Interleaving} 
From a technical point of view, different aspects of LLVM IR files have to be adjusted in order to generate a properly instrumented binary.

\begin{figure}
    \centering
    \includegraphics[width=\linewidth]{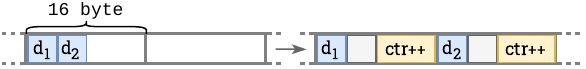}
    \caption{Interleaving counter values and small data chunks in memory. For freshness, the counter is incremented on each memory write. The data chunks are not packed to ease address calculations and elevate stability.}
    \label{fig:zebra-dummy}
\end{figure}

\paragraph{Interleaving Building Blocks}

First, for having access to a global counter in all contexts of the program, a vector register reserved for the counter is initialized with a random counter value generated by \texttt{rdrand}.
For easing the handling of global variables, we also add an initializer counter with a value that is known at compile time.
This initializer gets replaced with the vectorized counter during the first memory write to the global variable.
Similar to the counter, as a basic part of the mitigation, we also need to add all needed interleaved structs to insert proper handling for each primitive data type (e.g., integers with varying bit width, see~\Cref{fig:zebra-dummy}).

\begin{figure}
    \centering
    \includegraphics[width=0.9\linewidth]{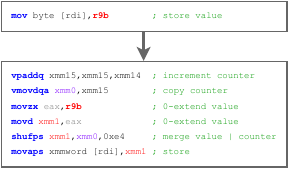}
    \caption{Instrumentation of a 1-byte store instruction. The value to be written is combined with a counter via register-only operations. A single instruction writes both value and counter.}
    \label{fig:store-asm}
\end{figure}

\paragraph{Memory-Based Instruction Adjustments}
Functions to be instrumented are determined by propagating a \texttt{clang} function attribute of a base function along the call graph of that function. 
Afterward, the call to the base function is updated to now call its instrumented equivalent that only has instrumented child functions.
Inside those instrumented functions, all address computations (called \texttt{GEP}s in LLVM), memory loads and stores (alongside instructions that contain implicit loads and stores like constant expression \texttt{GEP}s) and allocations have to be instrumented to follow the layout of interleaved and thereby extended memory.
Global variables and named structs also have to be adjusted to an interleaved format.
An example of the instrumentation of a memory write including the counter update can be found in~\Cref{fig:store-asm}.

\begin{figure}
    \centering
    \includegraphics[width=0.9\linewidth]{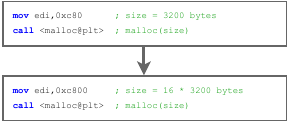}
    \caption{Instrumentation of a \texttt{malloc} call that allocates a byte array. The allocation size is adjusted by the compiler to fit the additional counters and padding.}
    \label{fig:malloc-asm}
\end{figure}

\paragraph{Heap Allocation Handling}
A share of the memory being extended by adding interleaving are heap allocations.
In order to ensure enough available space for adding interleaved structs on the heap, we extend the memory allocated by calls to \texttt{malloc} and related allocation functions.
As \texttt{malloc} does not use type information, we assume the ``worst'' case interleaved struct that contains one byte of data in each 16-byte block. 
An example of an instrumented version of \texttt{malloc} is given in~\Cref{fig:malloc-asm}.
We do not add the same SVF pointer analysis pass for type inference we added for \texttt{memcpy} or \texttt{memset} to lower the amount of manual intervention needed during compilation for fixing unknown cases. 
If type inference is needed to reduce memory overhead for certain examples, this could be adjusted.
While compiler intrinsics for copying or setting values can break with a default handling, the handling of heap allocations works stable when setting the default to the byte-wise handling.
Combined with the extension of the allocated heap space, the writes to heap objects are instrumented to not only include the writes of original data but also the 16-byte writes that combine original data and an incremented counter.

\subsection{Binary-Level Assessment}
\label{subsec:memtracer}
To ensure that the LLVM middle end level instrumentation is not removed by compiler passes that occur later in the compilation process, and that inserted stack spills do not leak any patterns of repeating values, we implement a binary-level assessment tool.
The assessment tool is a Pintool~\cite{DBLP:conf/pldi/LukCMPKLWRH05} that traces all memory writes of the instrumented binary and manages a set of all written values per address.
Whenever there is a repetition of written values within a set, the corresponding instruction offsets and written values are logged to a trace file so that the developer can manually check them for potential leakages in accordance with known leaking algorithms.
That way, the developer can determine whether secret data would be spilled to the stack and can apply some source-level mitigations like storing the affected data as global variables instead.
Afterward, the compilation and binary-level check is repeated until a non-leaky version is produced (cf.~\Cref{fig:zebra-framework}).

\section{\tool{} Evaluation}
\label{sec:eval}

In order to categorize the presented approach of \tool{} as a mitigation against \ctsc{} leakage, we assess the security and performance in comparison to the \cf{} masking approach.
We perform our evaluation on a system with an Intel 4th Gen Xeon Gold 6438Y+ CPU (Sapphire Rapids) with \num{512} GiB of DDR5 and local SSD storage. 
We base our evaluation on target primitives from \lsodium{}-1.0.20 and \mbedtls{}-3.6.0.

\subsection{Compile Time and Memory Consumption} 
To estimate the additional memory consumption and the additional time required to run the instrumentation process, we base our evaluation on the reference implementations of primitives in \lsodium{} and \mbedtls{}.
The reference implementations conform to~\Cref{subsec:limitations}, i.e., we preempt target-specific primitive versions that include inline assembly, which is out of scope for compiler-level analysis.
We have disabled vectorization flags for building the merged bitcode file that is instrumented by the \tool{} LLVM passes.
While a production-level version of the code may have less overhead in code size due to optimized inline assembly and less compilation time due to parallel build processes, we can better estimate the additional overhead.
The library build process happens only once, and we verify that the order of magnitude of the build remains the same.
To measure the runtime overhead (which is present within each execution), we compare to a vectorized baseline.

The compile time is composed of different compilation steps.
For measurements of compile time and memory consumption, we first build a non-instrumented version without vectorization of the library composed of various bitcode files.
We measure the time it takes to combine the necessary bitcode files into a non-instrumented version as well as into an instrumented version.
The compile time that is needed for building instrumented binaries using a pre-built library runs up to eight seconds while the sequential non-instrumented build process takes around three seconds.

The average code size of instrumented binaries increases by around \num{63}\% (from \num{220}/\num{250} MB to approx. \num{290}/\num{435} MB), where a recurring part is counter management and the rest belongs to other adjustments of instructions and global variables. 
The average memory usage increases by a factor of five; this is mostly due to lack of optimizations like checking the heap data types or filling each 8-byte data chunk of the interleaved structs with data. 
The comparatively high memory usage increase in \sha{} is an example of such missing heap data type optimizations.
More details can be found in~\Cref{tab:eval-general}.

\begin{table}[]
    \caption{Overhead of compilation time, code size and memory consumption. Compilation time and code size are related to the library version containing certain instrumented primitives (baseline given in the library rows), whereas memory usage is estimated for the execution of the primitive in isolation.}
    \label{tab:eval-general}
    \centering
    \small
    \begin{tabular}{ l c c r r } 
    \toprule
    \multirow{2}[3]{*}{\hspace*{-0.5em}Target} & \multirow{2}[3]{*}{\hspace*{-1em}\shortstack{Comp. time\\ (s)}} & \multirow{2}[3]{*}{\hspace*{-0.5em}\shortstack{Code size\\ (MB)}} & \multicolumn{2}{c}{Memory usage (KB)} \\
    \cmidrule(lr){4-5}
      &  &  & original & instrum. \\
    \midrule
    \multicolumn{1}{l}{\hspace*{-0.5em}\underline{\lsodium{}}} & \num{3.4} & \num{250}   & &\\
    \eddsa{} & \num{8.1} & \num{437}  & \num{2744} & \num{10336} \\
    \sha{} & \num{8.3} & \num{454}  & \num{5600} & \num{59568} \\
    \chacha{} & \num{7.6} & \num{389}  & \num{2288} & \num{7488} \\
    \midrule
    \multicolumn{1}{l}{\hspace*{-0.5em}\underline{\mbedtls{}}}    & \num{2.6} & \num{218} &  &  \\
    \baseLXIV{} & \num{5.8} & \num{286} & \num{1544} & \num{3488} \\
    \bottomrule
    \end{tabular}
\end{table}

\subsection{Runtime Overhead and Security}
\label{subsec:zebrafix-perf}
As \tool{} provides nonce-based randomization within the selected block size, the resulting encryption is fully probabilistic (analogous to CTR-mode encryption). 
With a block size of 16 bytes, of which 8 byte are counters, a repetition occurs only after $2^{64}$ write accesses.
In our implementation, every memory write is protected.
The exception to this are stack spills that are introduced after the LLVM IR instrumentation because of too much register pressure in the vector registers for storing intermediate variables.
The stack spills are displayed to the developer during the compile process and can further be checked at the binary-level (cf.~\Cref{subsec:memtracer}).

The primitives for the evaluation were chosen according to the following criteria: 
\texttt{EdDSA} has been shown to be vulnerable against \ctsc{} attacks~\cite{DBLP:conf/sp/LiWW0TZ22, DBLP:conf/uss/WichelmannPW023}. 
Apart from that, we include \baseLXIV{} as it was the fastest target in prior \ctsc{} mitigation evaluations, and \sha{} as another fast target for comparison.
We depict the \baseLXIV{} decoding in~\Cref{tab:eval}; the results for the encoding are similar.
All targets include zeroing of the resp. memory when freeing resources to avoid leakage of data through memory reuse.
As mentioned in~\Cref{subsec:limitations}, we exclude target versions that contain inline assembly or features that the current proof-of-concept does not support, like \texttt{AES} vector instructions.
In the latter case, the computation mainly happens in registers, so rare leaking spots can be subject to manual adjustments.

We measured the runtime overhead of \tool{} for each primitive in relation to an execution of the baseline version that is using a plain LLVM version.
The baseline version and the instrumented version share the same build parameters where possible (standard parameters with additional \emph{-{}-disable-asm}) and include disabling the vectorization for the instrumented version to ensure that all instructions can be properly instrumented by \tool{}.
The \tool{} instrumentation adds an average overhead factor of $1.3\times$ over all targets, whereas the most balanced version of \cf{} adds an average overhead of $4.2\times$ and the most secure version an overhead of $21.8\times$. 
The detailed results are shown in~\Cref{tab:eval}.

Although the memory tracking Pintool described in~\Cref{subsec:memtracer} does not provide information on the exploitability of potential leakage, code can be divided into known vulnerable functions where developers need to pay more attention during manual analyses, and other functions containing colliding written values.
Filtering the instruction offsets has verified that the known vulnerable functions do not contain any leaky memory writes.
Other leakages that need to be checked in other functions are \texttt{push} instructions and about 25 unprotected memory writes in the described targets that are not interleaved 16-byte writes.

\subsection{Comparison to \cf{} Masking Approach} 
\tool{} and \cf{} introduce memory overhead from the \ctsc{} protection compared to baseline execution.
While \tool{} extends each data chunk with a counter in interleaved memory writes, \cf{} is based on masking each plaintext chunk via XORing the data and the random mask.
In \cf{}, the \emph{Base} and \emph{Enhanced} versions contain a same-sized secrecy and mask buffer for each data chunk.
In the \cf{}-\emph{Fast} version, there is only one merged buffer for each data chunk.
Thereby, the memory overhead in \tool{} is equal to \cf{} or even smaller if 8-byte chunks are interleaved, whereas \tool{} introduces a larger memory overhead for smaller data types (\Cref{fig:zebra-dummy}).

\paragraph{Security}
From the different levels of security in \cf{}, \cf{}-Enhanced with \texttt{rdrand} is closest to the guarantees \tool{} provides in the \ctsc{} protection scenario.
This is due to the fact that masking requires adequate randomness in order to counteract the unpredictability of the data, and to ensure that the mask and the combined changes in the data do not cancel each other out.
An example of such a cancellation would be a leaky implementation that always writes either 0 or 1 to an address in combination with a mask that is only incremented on each write: Each time the mask changes the parity from even to odd, the last bit of the incrementing mask changes. 
If the written value is 0, the change of the mask bits results in a new plaintext, while the XOR of data and mask is a repetition of the prior value if a 1 is to be written.
Of all \cf{} versions, \cf{}-Base with the \xsp{} (\texttt{XS+}) randomness generator~\cite{DBLP:journals/jcam/Vigna17} provides the most balanced tradeoff between security and performance in the masking approach whereas \cf{}-Fast with \texttt{AES} has the weakest security guarantees but best performance.
In the case of stack spills, the \tool{} instrumentation security level might be collated with \cf{}-Base with \texttt{XS+} as there might be cases of repeating memory writes of secret data that have to be checked manually. 

\paragraph{Performance Comparison}
Even for the fastest primitive with the weakest security level in the \cf{} evaluation, \baseLXIV{} from \mbedtls{} and \cf{}-Fast with \texttt{AES}, \tool{} reaches a similar overhead. 
In \cf{}, only secret-dependent instructions are protected.
Thus, \tool{}'s performance impact could be reduced even further with secret tracking and other optimizations, e.g., allowing the use of vector instructions combined with interleaved memory writes.

In general, our results suggest that our proposed interleaving approach is more efficient than the data masking approach taken by \cf.
This is due to the fact that randomness generation introduces a significant overhead that we avoid. 
The remaining speedup of \tool{} over \cf{} can be explained with a more optimized compiler-built binary, instead of relying on binary rewriting.
We conclude that implementing an interleaving-based \ctsc{} mitigation in the compiler middle end can enhance performance and security compared to masking-based binary-level approaches.

\begin{table}[]
    \caption{Performance comparison to \cf{}. Values are given for \cf{}-Enhanced with \texttt{rdrand} (CF-E), \cf{}-Fast with \texttt{AES} (CF-F), \cf{}-Base with \texttt{XS+} (CF-B), and \tool{}. The numbers for \cf{} are based on results reported in their paper.}
    \label{tab:eval}
    \centering
    \small
    \begin{tabular}{l rrrr}
        \toprule
        Target & CF-E & CF-F & CF-B & \tool{} \\
         \midrule 
         \multicolumn{5}{l}{\hspace*{-0.5em}\underline{\lsodium{}}}\\
         \eddsa{} & \num{39.1}$\times$ & \num{5.5}$\times$ & \num{8.6}$\times$ & \num{1.6}$\times$ \\
         \sha{} & \num{21.6}$\times$ & \num{1.6}$\times$ & \num{2.4}$\times$ & \num{1.3}$\times$ \\ 
         \midrule
         \multicolumn{5}{l}{\hspace*{-0.5em}\underline{\mbedtls{}}}\\
         \baseLXIV{} & \num{4.6}$\times$ & \num{1.2}$\times$ & \num{1.6}$\times$ & \num{1.1}$\times$ \\
         \bottomrule
    \end{tabular}
\end{table}

\section{Preventing Leakages from \SsssTitle{}}
\label{sec:prev-sss}

With the discussion of \ctsc{} defenses in mind, we now explore connections to \ssss{} to show how these \scname{} leakages can be generalized to more holistic defenses.
Per se, all attacks described in~\Cref{sec:background} are based on different mechanisms to exploit leakage.
However, despite originating from different sources, all of these leakages can be considered \scname{} side-channels and thus be thwarted by principled defenses:
If the data was not architecturally written to memory in a leaky way in the first place, the different leakage mechanisms are no longer effective.
In this section, we outline the interconnection of \ctsc{} and \sss{} leakages and compare the effectiveness of \tool{} with \sss{} mitigations.

\paragraph{Refined Threat Model} 
In the \ctsc{} leakage scenario, an attacker can observe deterministic patterns of ciphertext changes of victim VM data and thereby derive information about secret keys.
The attacker model of \ctsc{} leakage is the strongest one to be considered.
In the other \scname{} leakage scenarios, we assume that \uarchopts{} that are currently on the rise, like the ones discussed in~\Cref{sec:background}, will gain more importance and see implementation in common processors.
The attacker is able to bring the system into a state where the best conditions for maximal leakage through \uarchopts{} have been established; thus we assume the worst case leakage in each channel to be possible.\footnote{At the time of writing, we are not aware of systems that already feature the \uarchopts{} in question, apart from the ones that are linked in~\Cref{sec:background}.}
The assumptions on code that is instrumented are in line with the ones presented in~\Cref{subsec:threatmodel}.

\subsection{Interconnecting \SsssTitle{} and \CtscTitle{} Leakage}
\label{subsec:relation-ctsc-sss}
In \ch{}~\cite{DBLP:conf/uss/DengLTWYZ23}, the authors propose to use symbolic execution to detect \ctsc{} leakage.
Consecutive secret-dependent memory writes are checked for whether the values to be written are always the same, always different or possibly different.
Thereby, \ctsc{} leakage is found for the case of two consecutive memory writes of the same secret-dependent variable to the same address if the value of the variable may vary. 
However, the leakage model does not cover all \ctsc{} leakages. 
Missing cases include interleaved memory writes of secret values to the same address or leakage through observations of ciphertext changes between a known public value and a secret value (e.g., replacing an all-zero public value with a secret that can either be zero as well or some other value).
Thus, \ch{} only finds a case of \sss{} leakage, where two consecutive secret-dependent memory writes with possibly different values occur.

The \sss{} mitigation \cio{}~\cite{DBLP:conf/asplos/FlandersSMGK24} adds a memory write with a provably different value between two consecutive writes. 
Thereby, the \sss{} leakage is thwarted.
The proposed mitigation still allows observing the ciphertext patterns that lead to \ctsc{} leakage as the attacker just has to monitor more memory writes and remove the dummy writes (the ones that hinder \ssss{}) in between.
Even worse, the incomplete modeling of \ctsc{} leakage in \ch{} would incorrectly state that code instrumented with a \cio{}-style approach does not leak in the \ctsc{} scenario. 

However, any \ctsc{} defense also protects against \sss{} leakage if the option of silencing the store is checked for the same chunk width as the inserted memory write width.
In the case of rotating the location a data element is written to, and under the assumption that the memory is initialized properly, the memory addresses are not reused for the same data and are thereby protected against leakage from \ssss{}.
In the case of a probabilistic software encryption layer through masking or interleaving, all written values are different due to the probabilistic encryption, and there are no stores that can deterministically be suppressed by \uarchopts{}.

\subsection{Comparison of \SssTitle{} Mitigations}

The \tool{} implementation protects against leakage from \ssss{} without requiring further adaptions as long as the silencing of writes cannot be decomposed into smaller chunks.
We therefore compare \tool{} to the \sss{} mitigation \cio{}~\cite{DBLP:conf/asplos/FlandersSMGK24} that has also been implemented for x86-64.
The \tool{} implementation addresses the general root cause of \scname{} leakages.
In line with that, the proof-of-concept implementation does not include a tracking for leaking or secret-dependent instructions, but rather instruments all memory writes. 
The \sss{}-only version of \cio{} only covers a subset of all writes that are instrumented in \tool{}.
\tool{} does, in contrast to \cio{}, not have an automated verification mechanism but rather introduces manual assessment of the non-leakiness of memory writes that are introduced by the register allocation.

Out of the targets evaluated in \cio{}, we compare against \eddsa{} and \chacha{} which were shown to be (partially) vulnerable in~\cite{DBLP:conf/ccs/BartheBCCGGRSWY24}.
For \chacha{}, we refer to the encryption; the decryption results are similar.
The remaining targets evaluated in \cio{} are not included in this case study because they contain instructions that are not supported in our proof-of-concept due to the way the interleaving is implemented based on vector register writes.  
Yet, the analyzed primitives are implemented to be executed in vector registers so that the leaking points are initialization and memory cleanup.
It is possible to manually safeguard this limited set of code.

\paragraph{Performance Comparison}
The performance results for \cio{} are based on results reported in their paper.
The \tool{} measurement setup equals the one described in~\Cref{subsec:zebrafix-perf}.
In line with \cio{}, we use the \lsodium{} reference implementations and protect stores of up to 8 byte of data.  
Protecting a binary against \sss{} leakage with \cio{} introduces an overhead of \num{1.85}$\times$ for \chacha{} and an overhead of \num{3.76}$\times$ for \eddsa{}. 
In contrast to that, the \ctsc{} protection from \tool{} introduces an overhead factor of \num{2.1}$\times$ for \chacha{} and a factor of \num{1.6}$\times$ for \eddsa{}. 

Even without further secret tracking in \tool{} and including the fact that \ssss{} are only a subset of leaky memory writes in the \ctsc{} leakage model, \tool{} is faster than \cio{} in the case of \eddsa{}.
Other potentially vulnerable algorithms like \chacha{} contain more memory writes that are verified to be non-leaky in the case of \cio{} so that the relative overhead of execution to the baseline is increased when instrumenting with \tool{}.

\paragraph{Security}
The memory tracking Pintool without further adaption gives an overapproximation of the leakages through \ssss{} because if we can rule out leakages through any collision, we can also rule out leakages through consecutive memory writes leading to a collision.
The overapproximation can be refined if the Pintool is often needed for assessing implementations in the \sss{} scenario.

In line with~\cite{DBLP:conf/ccs/BartheBCCGGRSWY24}, we found certain possible leakages in the \chacha{} target.
However, those leakages are not originated in multiple systematic collisions as in \texttt{cmov} or \texttt{cswap} functionality but rather single collisions with a large set of possible values that are written during each execution.
According to current knowledge, we do not deem those writes as exploitable which is also in line with \cio{} being able to prune a large set of consecutive memory writes from their instrumentation.

Additionally, stack spills are introduced for the core \chacha{} functionality because the data in the \tool{} proof-of-concept implementation is not packed (see~\Cref{fig:zebra-dummy}). 
The amount of data that can fit into vector registers instead of being written to the stack could be improved to enhance both security guarantees and performance.
However, such an improvement would need precise and complete tracking of pointers to guarantee stability and functional correctness of the instrumented program.

\section{Lessons Learned for Future Mitigations}
\label{subsec:limitations}
In the following, we discuss the practical implications of mitigation implementations that aim for extensibility and generalizability, choosing to minimize architecture-specific tweaking where possible.

\paragraph{Feature Support}
As described in~\Cref{sec:leakdeflevels}, instrumenting on compiler middle end level has multiple advantages and disadvantages.
One of the main disadvantages of any static instrumentation method is that language features, such as determining the size of a buffer by subtracting pointers without knowing the content type, cannot be instrumented semantically correctly, and must be manually replaced in the source code.
Another common issue during our implementation was the evaluation of \texttt{sizeof}, which already happens in the compiler front end.
This occurred in cases where \texttt{sizeof} was used to determine the input size of operations like \texttt{memcpy} or \texttt{memset}, which work on arrays.
In our proof-of-concept implementation of \tool{}, we work around that problem by retrieving the correct byte count from the LLVM \texttt{DataLayout} object and combining that with type information from the SVF pass.

We support memory writes up to 8 bytes in size, which is the integer size on x86-64. 
For larger writes (e.g., vector instructions), the data would have to be split into chunks and written within multiple interleaved chunks.
Although we have not implemented vector support in \tool{}, we do not expect any specific performance impact when rewriting vectorized stores.
In contrast to the typical dynamic linking of cryptographic libraries, the interleaving-based leakage mitigation requires that either all dependencies are recompiled and statically linked or that the data from the instrumented code is properly handed over to unprotected functionality after declassification.
The extensive changes to the memory layout cannot be handled transparently by non-instrumented code.

\paragraph{Expert Knowledge} 
Most cryptographic libraries are implemented in C, a language that lacks type safety, which poses certain key challenges for memory layout adjustments.
Some manual intervention is needed for functions that use ``indirect'' casts on input data like putting a char array into a function that expects to work with 4-byte integers.
Such casts might lead to semantically wrong results when applying operations that depend on the data type's width like copying data from one buffer into another.
The default handling case can only assume 8-byte data chunks in order to avoid out-of-bounds reads or writes, so any function that displays a warning has to be checked and then adjusted in the LLVM SVF pass handling for the correct pointer tracking.
The currently implemented solution relies on the fact that function names are available in such cases for easing manual adjustments.
With C, there is the choice between manual code adaptation, significant compiler code analysis, or extensive runtime type tracking.
Alternatively, one can rely on a language in which such casts do not occur, and binary types of variables are always known.

\paragraph{General Limitations of Middle End Instrumentation}

A general issue with instrumentation at the LLVM IR level is that, despite offering architecture-independent behavior, it introduces a reliance on the compiler infrastructure to handle operations in a specific way.
One specific part that has to be re-checked with compiler updates are the vector writes that are introduced in IR.
These writes must be translated into equivalent 16-byte store instructions, and not broken up into smaller written operations.
Even though we did not insert intrinsics for the 16-byte writes, the default IR vector write handling in LLVM already successfully translated our interleaved writes correctly.

Another general issue arising in the proposed instrumentation approach stems from global variable usage: Either all code that uses global variables is instrumented or no code at all.
In case a mitigation is implemented with additional secrecy tracking, the correct functionality cannot be guaranteed for global variable usage.
If instrumented and non-instrumented parts of the code used global variables, the global variable state might destroy synchronization of values between different functions.
Moreover, the general functionality can be compromised as even if all global variable usages were replaced with the instrumented version, there are usages depending on the global variables that cause problems in function parameters or functions like \texttt{memcpy} or \texttt{memset} or in \texttt{GEP} accesses for address calculations.\footnote{Even with optimization level \texttt{-O2} and having all GEP accesses combined in one instruction due to the LLVM \emph{InstCombine} pass, the iteration steps in arrays can still be wrong if the source element type of a GEP is not interleaved while the pointer of it points to an interleaved (instrumented) global variable.} 
Further analyses would incur large overheads and make the resulting binary less stable; Apart from that, we deem the scenario of needing a non-instrumented version of each function alongside of instrumented ones unlikely and therefore instrument the whole bitcode file.

\section{Related Work}
\label{sec:related-work}
Much work has been put into analysis of programs for constant-time properties~\cite{DBLP:conf/uss/AlmeidaBBDE16, DBLP:conf/ccs/WichelmannSP022, DBLP:journals/tissec/DanielBR22}.
While there are still many open questions left~\cite{DBLP:conf/ccs/GeimerVRDBM23}, we assume programs that are in line with the constant-time paradigm.
Under the assumption of being constant-time, further leakages like \ssss{} of secret-dependent memory writes~\cite{DBLP:conf/uss/DengLTWYZ23} can be pinpointed.

Apart from analyzing programs for timing side-channel leakage, there are also leakage evaluations for examining leakage properties originating from upcoming \uarchopts{}~\cite{DBLP:conf/isca/VicarteSNT0KF21, DBLP:conf/ccs/BartheBCCGGRSWY24}.
With the possibility of pinpointing leaking code pieces, many leakage mitigations approaches are proposed:
Obelix~\cite{DBLP:conf/sp/WichelmannRPE24} is an ORAM-based tool that combines many side-channel leakage mitigations.
Obelix also provides a protection level that mitigates \ctsc{} leakage through data rotation and interleaving but produces overhead factors around $1000\times$ and does not scale to larger applications.
In case only specific features are supported by the architecture, mitigations introducing lower overhead can be applied~\cite{DBLP:conf/eurosp/WinderixMP21}; for open architectures even with instruction set architecture (ISA) augmentations~\cite{DBLP:conf/eurosp/BognarWBP23, DBLP:conf/sp/WinderixBNDP24}.
However, we assume a cloud setting where we cannot rely on ISA augmentations, making the proposed approaches inapplicable.

By enabling hardware configurations that disable certain optimizations, certain side-channel leakages can be avoided.
For operand-independent timing of instructions, ARM introduced DIT~\cite{arm-dit} and Intel DOIT~\cite{intel-doit}.
The operand-independent timing modes disable \dmping{}~\cite{intelDdp,DBLP:conf/uss/ChenWSFKPG24}.
Yet, the actual implementation is unclear and if the hypervisor can adjust the modes, software solutions are still needed as a backup.
A discussion of potential software-based solutions is given in~\Cref{app:dmp}.

An extension of DSR is CoDaRR where the data gets re-randomized in tight intervals~\cite{DBLP:conf/ccs/RajasekaranCGNV20}.
However, the overhead from such a re-randomization for all data does not scale for \ctsc{} protection with recurring memory writes.
Schemes like Data and Pointer Prioritization (DPP)~\cite{DBLP:conf/uss/0001LJHAY23} that try optimizing similar overheads scale well only when their protection is restricted to data that might be subject to memory errors and associated unauthorized memory writes.
Works like~\cite{DBLP:conf/acsac/PalitMP19} and~\cite{DBLP:conf/sp/PalitMMP21} promise selective encryption for data in memory.
Yet, the encryption mode does not include freshness and thus does not add protection for multiple memory writes with the same value to the same address as needed for \ctsc{} mitigation; adding key updates introduces the common problem of randomness generation.

\section{Conclusion}
\label{sec:conclusion}

In this work, we have studied \scname{} side-channel leakages that render constant-time code guarantees ineffective.
We have discussed prerequisites for efficient \ctsc{} mitigations and implemented \tool{}.
We have shown that interleaving can improve both the performance and security guarantees of \ctsc{} mitigations, though this may come with practical constraints such as source code adjustments.
Moreover, we have leveraged interleaving to devise mitigations against more generic \scname{} side-channel leakages like \ssss{}, and explained how holistic defenses can be developed.
We conclude that, despite its limitations, interleaving is a promising approach to counter \ctscs{} and \ssss{}.

\section*{Acknowledgment}
We would like to thank the anonymous reviewers for their thorough assessment and providing helpful suggestions for improvement.
We also thank Johannes Liebenow and Pajam Pauls for proofreading and valuable feedback.
This work has been supported by Deutsche Forschungsgemeinschaft (DFG) through the ReTEE project, and by Bundesministerium für Bildung und Forschung (BMBF) through the ENCOPIA project.

\bibliographystyle{IEEEtranS}{%
    \footnotesize%
    \bibliography{ref}%

\begin{thebibliography}{10}
\providecommand{\url}[1]{#1}
\csname url@samestyle\endcsname
\providecommand{\newblock}{\relax}
\providecommand{\bibinfo}[2]{#2}
\providecommand{\BIBentrySTDinterwordspacing}{\spaceskip=0pt\relax}
\providecommand{\BIBentryALTinterwordstretchfactor}{4}
\providecommand{\BIBentryALTinterwordspacing}{\spaceskip=\fontdimen2\font plus
\BIBentryALTinterwordstretchfactor\fontdimen3\font minus \fontdimen4\font\relax}
\providecommand{\BIBforeignlanguage}[2]{{%
\expandafter\ifx\csname l@#1\endcsname\relax
\typeout{** WARNING: IEEEtranS.bst: No hyphenation pattern has been}%
\typeout{** loaded for the language `#1'. Using the pattern for}%
\typeout{** the default language instead.}%
\else
\language=\csname l@#1\endcsname
\fi
#2}}
\providecommand{\BIBdecl}{\relax}
\BIBdecl

\bibitem{DBLP:conf/uss/0001LJHAY23}
\BIBentryALTinterwordspacing
S.~Ahmed, H.~Liljestrand, H.~Jamjoom, M.~Hicks, N.~Asokan, and D.~Yao, ``{Not All Data are Created Equal: Data and Pointer Prioritization for Scalable Protection Against Data-Oriented Attacks},'' in \emph{32nd {USENIX} Security Symposium}.\hskip 1em plus 0.5em minus 0.4em\relax {USENIX} Association, 2023. [Online]. Available: \url{https://www.usenix.org/conference/usenixsecurity23/presentation/ahmed-salman}
\BIBentrySTDinterwordspacing

\bibitem{DBLP:conf/uss/AlmeidaBBDE16}
\BIBentryALTinterwordspacing
J.~B. Almeida, M.~Barbosa, G.~Barthe, F.~Dupressoir, and M.~Emmi, ``{Verifying Constant-Time Implementations},'' in \emph{25th {USENIX} Security Symposium}, T.~Holz and S.~Savage, Eds.\hskip 1em plus 0.5em minus 0.4em\relax {USENIX} Association, 2016, pp. 53--70. [Online]. Available: \url{https://www.usenix.org/conference/usenixsecurity16/technical-sessions/presentation/almeida}
\BIBentrySTDinterwordspacing

\bibitem{SEV-SNP_whitepaper}
\BIBentryALTinterwordspacing
AMD, ``{AMD SEV-SNP: Strengthening VM Isolation with Integrity Protection and More}.'' [Online]. Available: \url{https://www.amd.com/system/files/TechDocs/SEV-SNP-strengthening-vm-isolation-with-integrity-protection-and-more.pdf}
\BIBentrySTDinterwordspacing

\bibitem{AMD-ciphertext-hiding}
\BIBentryALTinterwordspacing
{AMD}, ``{SEV Secure Nested Paging Firmware ABI Specification}.'' [Online]. Available: \url{https://www.amd.com/content/dam/amd/en/documents/epyc-technical-docs/specifications/56860.pdf}
\BIBentrySTDinterwordspacing

\bibitem{andriesse2018practical}
D.~Andriesse, \emph{{Practical Binary Analysis: Build Your Own Linux Tools for Binary Instrumentation, Analysis, and Disassembly}}.\hskip 1em plus 0.5em minus 0.4em\relax No Starch Press, 2018.

\bibitem{arm-dit}
\BIBentryALTinterwordspacing
ARM, ``{DIT: Data Independent Timing}.'' [Online]. Available: \url{https://developer.arm.com/documentation/ddi0601/2024-06/AArch64-Registers/DIT--Data-Independent-Timing}
\BIBentrySTDinterwordspacing

\bibitem{DBLP:conf/ccs/BartheBCCGGRSWY24}
\BIBentryALTinterwordspacing
G.~Barthe, M.~B{\"{o}}hme, S.~Cauligi, C.~Chuengsatiansup, D.~Genkin, M.~Guarnieri, D.~M. Romero, P.~Schwabe, D.~Wu, and Y.~Yarom, ``{Testing Side-Channel Security of Cryptographic Implementations Against Future Microarchitectures},'' in \emph{2024 {ACM} {SIGSAC} Conference on Computer and Communications Security ({CCS})}.\hskip 1em plus 0.5em minus 0.4em\relax {ACM}, 2024, pp. 1076--1090. [Online]. Available: \url{https://doi.org/10.1145/3658644.3670319}
\BIBentrySTDinterwordspacing

\bibitem{DBLP:conf/uss/BartolomeoMP23}
\BIBentryALTinterwordspacing
L.~D. Bartolomeo, H.~Moghaddas, and M.~Payer, ``{ARMore: Pushing Love Back Into Binaries},'' in \emph{32nd {USENIX} Security Symposium}, J.~A. Calandrino and C.~Troncoso, Eds.\hskip 1em plus 0.5em minus 0.4em\relax {USENIX} Association, 2023, pp. 6311--6328. [Online]. Available: \url{https://www.usenix.org/conference/usenixsecurity23/presentation/di-bartolomeo}
\BIBentrySTDinterwordspacing

\bibitem{DBLP:conf/eurosp/BognarWBP23}
M.~Bognar, H.~Winderix, J.~V. Bulck, and F.~Piessens, ``{MicroProfiler: Principled Side-Channel Mitigation through Microarchitectural Profiling},'' in \emph{8th {IEEE} European Symposium on Security and Privacy (EuroS{\&}P)}.\hskip 1em plus 0.5em minus 0.4em\relax {IEEE}, 2023, pp. 651--670.

\bibitem{DBLP:conf/uss/ChenWSFKPG24}
\BIBentryALTinterwordspacing
B.~Chen, Y.~Wang, P.~Shome, C.~W. Fletcher, D.~Kohlbrenner, R.~Paccagnella, and D.~Genkin, ``{GoFetch: Breaking Constant-Time Cryptographic Implementations Using Data Memory-Dependent Prefetchers},'' in \emph{33rd {USENIX} Security Symposium}.\hskip 1em plus 0.5em minus 0.4em\relax {USENIX} Association, 2024. [Online]. Available: \url{https://www.usenix.org/conference/usenixsecurity24/presentation/chen-boru}
\BIBentrySTDinterwordspacing

\bibitem{DBLP:journals/tissec/DanielBR22}
L.~Daniel, S.~Bardin, and T.~Rezk, ``{Binsec/Rel: Symbolic Binary Analyzer for Security with Applications to Constant-Time and Secret-Erasure},'' \emph{{ACM} Trans. Priv. Secur.}, vol.~26, no.~2, pp. 11:1--11:42, 2023.

\bibitem{DBLP:conf/uss/DengLTWYZ23}
\BIBentryALTinterwordspacing
S.~Deng, M.~Li, Y.~Tang, S.~Wang, S.~Yan, and Y.~Zhang, ``{CipherH: Automated Detection of Ciphertext Side-channel Vulnerabilities in Cryptographic Implementations},'' in \emph{32nd {USENIX} Security Symposium}, J.~A. Calandrino and C.~Troncoso, Eds.\hskip 1em plus 0.5em minus 0.4em\relax {USENIX} Association, 2023, pp. 6843--6860. [Online]. Available: \url{https://www.usenix.org/conference/usenixsecurity23/presentation/deng-sen}
\BIBentrySTDinterwordspacing

\bibitem{DBLP:conf/sp/DineshBXP20}
S.~Dinesh, N.~Burow, D.~Xu, and M.~Payer, ``{RetroWrite: Statically Instrumenting COTS Binaries for Fuzzing and Sanitization},'' in \emph{2020 {IEEE} Symposium on Security and Privacy (S\&P)}.\hskip 1em plus 0.5em minus 0.4em\relax {IEEE}, 2020, pp. 1497--1511.

\bibitem{intelSilentStores}
\BIBentryALTinterwordspacing
T.~Downs, ``{Hardware Store Elimination}.'' [Online]. Available: \url{https://travisdowns.github.io/blog/2020/05/13/intel-zero-opt.html}
\BIBentrySTDinterwordspacing

\bibitem{DBLP:conf/asplos/FlandersSMGK24}
M.~Flanders, R.~K. Sharma, A.~E. Michael, D.~Grossman, and D.~Kohlbrenner, ``{Avoiding Instruction-Centric Microarchitectural Timing Channels Via Binary-Code Transformations},'' in \emph{2024 Architectural Support for Programming Languages and Operating Systems ({ASPLOS})}.\hskip 1em plus 0.5em minus 0.4em\relax {ACM}, 2024.

\bibitem{DBLP:conf/ccs/GeimerVRDBM23}
A.~Geimer, M.~Vergnolle, F.~Recoules, L.~Daniel, S.~Bardin, and C.~Maurice, ``{A Systematic Evaluation of Automated Tools for Side-Channel Vulnerabilities Detection in Cryptographic Libraries},'' in \emph{2023 {ACM} {SIGSAC} Conference on Computer and Communications Security ({CCS})}, W.~Meng, C.~D. Jensen, C.~Cremers, and E.~Kirda, Eds.\hskip 1em plus 0.5em minus 0.4em\relax {ACM}, 2023, pp. 1690--1704.

\bibitem{intelDdp}
\BIBentryALTinterwordspacing
Intel, ``{Data Dependent Prefetcher}.'' [Online]. Available: \url{https://www.intel.com/content/www/us/en/developer/articles/technical/software-security-guidance/technical-documentation/data-dependent-prefetcher.html}
\BIBentrySTDinterwordspacing

\bibitem{intel-doit}
\BIBentryALTinterwordspacing
{Intel}, ``{Data Operand Independent Timing Instruction Set Architecture (ISA) Guidance}.'' [Online]. Available: \url{https://www.intel.com/content/www/us/en/developer/articles/technical/software-security-guidance/best-practices/data-operand-independent-timing-isa-guidance.html}
\BIBentrySTDinterwordspacing

\bibitem{DBLP:conf/cgo/LattnerA04}
C.~Lattner and V.~S. Adve, ``{LLVM: A Compilation Framework for Lifelong Program Analysis \& Transformation},'' in \emph{2nd {IEEE} / {ACM} International Symposium on Code Generation and Optimization ({CGO})}.\hskip 1em plus 0.5em minus 0.4em\relax {IEEE} Computer Society, 2004, pp. 75--88.

\bibitem{DBLP:conf/isca/LepakL00}
K.~M. Lepak and M.~H. Lipasti, ``{On the Value Locality of Store Instructions},'' in \emph{{ACM/IEEE} 27th International Symposium on Computer Architecture ({ISCA})}.\hskip 1em plus 0.5em minus 0.4em\relax {IEEE} Computer Society, 2000.

\bibitem{DBLP:conf/sp/LiWW0TZ22}
M.~Li, L.~Wilke, J.~Wichelmann, T.~Eisenbarth, R.~Teodorescu, and Y.~Zhang, ``{A Systematic Look at Ciphertext Side Channels on AMD SEV-SNP},'' in \emph{2022 {IEEE} Symposium on Security and Privacy (S\&P)}.\hskip 1em plus 0.5em minus 0.4em\relax {IEEE}, 2022, pp. 337--351.

\bibitem{DBLP:conf/uss/LiZWLC21}
\BIBentryALTinterwordspacing
M.~Li, Y.~Zhang, H.~Wang, K.~Li, and Y.~Cheng, ``{CipherLeaks: Breaking Constant-time Cryptography on AMD SEV via the Ciphertext Side Channel},'' in \emph{30th {USENIX} Security Symposium}, M.~Bailey and R.~Greenstadt, Eds.\hskip 1em plus 0.5em minus 0.4em\relax {USENIX} Association, 2021, pp. 717--732. [Online]. Available: \url{https://www.usenix.org/conference/usenixsecurity21/presentation/li-mengyuan}
\BIBentrySTDinterwordspacing

\bibitem{DBLP:conf/pldi/LukCMPKLWRH05}
C.~Luk, R.~S. Cohn, R.~Muth, H.~Patil, A.~Klauser, P.~G. Lowney, S.~Wallace, V.~J. Reddi, and K.~M. Hazelwood, ``{Pin: Building Customized Program Analysis Tools with Dynamic Instrumentation},'' in \emph{2005 {ACM} {SIGPLAN} International Conference on Programming Language Design and Implementation ({PLDI})}, V.~Sarkar and M.~W. Hall, Eds.\hskip 1em plus 0.5em minus 0.4em\relax {ACM}, 2005, pp. 190--200.

\bibitem{DBLP:journals/ijpp/MoghimiWES19}
A.~Moghimi, J.~Wichelmann, T.~Eisenbarth, and B.~Sunar, ``{MemJam: A False Dependency Attack Against Constant-Time Crypto Implementations},'' \emph{{Int. J. Parallel Program.}}, vol.~47, no.~4, pp. 538--570, 2019.

\bibitem{DBLP:conf/acsac/PalitMP19}
T.~Palit, F.~Monrose, and M.~Polychronakis, ``{Mitigating Data Leakage by Protecting Memory-Resident Sensitive Data},'' in \emph{35th Annual Computer Security Applications Conference (ACSAC)}.\hskip 1em plus 0.5em minus 0.4em\relax {ACM}, 2019.

\bibitem{DBLP:conf/sp/PalitMMP21}
T.~Palit, J.~F. Moon, F.~Monrose, and M.~Polychronakis, ``{DynPTA: Combining Static and Dynamic Analysis for Practical Selective Data Protection},'' in \emph{2021 {IEEE} Symposium on Security and Privacy (S\&P)}.\hskip 1em plus 0.5em minus 0.4em\relax {IEEE}, 2021, pp. 1919--1937.

\bibitem{DBLP:conf/ccs/RajasekaranCGNV20}
P.~Rajasekaran, S.~Crane, D.~Gens, Y.~Na, S.~Volckaert, and M.~Franz, ``{CoDaRR: Continuous Data Space Randomization against Data-Only Attacks},'' in \emph{2020 {ACM} Asia Conference on Computer and Communications Security ({ASIA} {CCS})}, H.~Sun, S.~Shieh, G.~Gu, and G.~Ateniese, Eds.\hskip 1em plus 0.5em minus 0.4em\relax {ACM}, 2020, pp. 494--505.

\bibitem{DBLP:conf/ccs/SieckBW021}
F.~Sieck, S.~Berndt, J.~Wichelmann, and T.~Eisenbarth, ``{Util::Lookup: Exploiting Key Decoding in Cryptographic Libraries},'' in \emph{2021 {ACM} {SIGSAC} Conference on Computer and Communications Security ({CCS})}, Y.~Kim, J.~Kim, G.~Vigna, and E.~Shi, Eds.\hskip 1em plus 0.5em minus 0.4em\relax {ACM}, 2021, pp. 2456--2473.

\bibitem{DBLP:journals/tches/SieckZBCEY24}
F.~Sieck, Z.~Zhang, S.~Berndt, C.~Chuengsatiansup, T.~Eisenbarth, and Y.~Yarom, ``{TeeJam: Sub-Cache-Line Leakages Strike Back},'' \emph{{IACR} Trans. Cryptogr. Hardw. Embed. Syst.}, vol. 2024, no.~1, pp. 457--500, 2024.

\bibitem{DBLP:conf/cc/SuiX16}
Y.~Sui and J.~Xue, ``{SVF: Interprocedural Static Value-Flow Analysis in LLVM},'' in \emph{25th International Conference on Compiler Construction ({CC})}, A.~Zaks and M.~V. Hermenegildo, Eds.\hskip 1em plus 0.5em minus 0.4em\relax {ACM}, 2016.

\bibitem{DBLP:conf/sp/VicarteFPG0FK22}
J.~R.~S. Vicarte, M.~Flanders, R.~Paccagnella, G.~Garrett{-}Grossman, A.~Morrison, C.~W. Fletcher, and D.~Kohlbrenner, ``{Augury: Using Data Memory-Dependent Prefetchers to Leak Data at Rest},'' in \emph{2022 {IEEE} Symposium on Security and Privacy (S\&P)}.\hskip 1em plus 0.5em minus 0.4em\relax {IEEE}, 2022, pp. 1491--1505.

\bibitem{DBLP:conf/isca/VicarteSNT0KF21}
J.~R.~S. Vicarte, P.~Shome, N.~Nayak, C.~Trippel, A.~Morrison, D.~Kohlbrenner, and C.~W. Fletcher, ``{Opening Pandora's Box: A Systematic Study of New Ways Microarchitecture Can Leak Private Data},'' in \emph{48th {ACM/IEEE} Annual International Symposium on Computer Architecture ({ISCA})}.\hskip 1em plus 0.5em minus 0.4em\relax {IEEE}, 2021, pp. 347--360.

\bibitem{DBLP:journals/jcam/Vigna17}
S.~Vigna, ``{Further Scramblings of Marsaglia's Xorshift Generators},'' \emph{{J. Comput. Appl. Math.}}, vol. 315, pp. 175--181, 2017.

\bibitem{wang2025peekPreprint}
\BIBentryALTinterwordspacing
A.~Wang, B.~Chen, Y.~Wang, C.~Fletcher, D.~Genkin, D.~Kohlbrenner, and R.~Paccagnella, ``{Peek-a-Walk: Leaking Secrets via Page Walk Side Channels},'' in \emph{2025 {IEEE} Symposium on Security and Privacy (S\&P)}.\hskip 1em plus 0.5em minus 0.4em\relax {IEEE}, 2025. [Online]. Available: \url{https://www.computer.org/csdl/proceedings-article/sp/2025/223600a023/21B7QepK7Fm}
\BIBentrySTDinterwordspacing

\bibitem{riscvSilentStores}
\BIBentryALTinterwordspacing
A.~Waterman and K.~Asanovic, ``{The RISC-V Instruction Set Manual, Volume I: Unprivileged ISA (Document Version 20191213)}.'' [Online]. Available: \url{https://riscv.org/wp-content/uploads/2019/12/riscv-spec-20191213.pdf}
\BIBentrySTDinterwordspacing

\bibitem{cipherfixSlides}
\BIBentryALTinterwordspacing
J.~Wichelmann and A.~P{\"{a}}tschke, ``{Cipherfix: Mitigating Ciphertext Side-Channel Attacks in Software (Slides)}.'' [Online]. Available: \url{https://www.usenix.org/system/files/sec23_slides_wichelmann.pdf}
\BIBentrySTDinterwordspacing

\bibitem{DBLP:conf/uss/WichelmannPW023}
\BIBentryALTinterwordspacing
J.~Wichelmann, A.~P{\"{a}}tschke, L.~Wilke, and T.~Eisenbarth, ``{Cipherfix: Mitigating Ciphertext Side-Channel Attacks in Software},'' in \emph{32nd {USENIX} Security Symposium}, J.~A. Calandrino and C.~Troncoso, Eds.\hskip 1em plus 0.5em minus 0.4em\relax {USENIX} Association, 2023. [Online]. Available: \url{https://www.usenix.org/conference/usenixsecurity23/presentation/wichelmann}
\BIBentrySTDinterwordspacing

\bibitem{DBLP:conf/sp/WichelmannRPE24}
J.~Wichelmann, A.~Rabich, A.~P\"atschke, and T.~Eisenbarth, ``{Obelix: Mitigating Side-Channels through Dynamic Obfuscation},'' in \emph{2024 {IEEE} Symposium on Security and Privacy (S\&P)}.\hskip 1em plus 0.5em minus 0.4em\relax {IEEE}, 2024.

\bibitem{DBLP:conf/ccs/WichelmannSP022}
J.~Wichelmann, F.~Sieck, A.~P{\"{a}}tschke, and T.~Eisenbarth, ``{Microwalk-CI: Practical Side-Channel Analysis for JavaScript Applications},'' in \emph{2022 {ACM} {SIGSAC} Conference on Computer and Communications Security ({CCS})}, H.~Yin, A.~Stavrou, C.~Cremers, and E.~Shi, Eds.\hskip 1em plus 0.5em minus 0.4em\relax {ACM}, 2022, pp. 2915--2929.

\bibitem{DBLP:conf/sp/WinderixBNDP24}
\BIBentryALTinterwordspacing
H.~Winderix, M.~Bognar, J.~Noorman, L.-A. Daniel, and F.~Piessens, ``{Architectural Mimicry: Innovative Instructions to Efficiently Address Control-Flow Leakage in Data-Oblivious Programs},'' in \emph{2024 {IEEE} Symposium on Security and Privacy (S\&P)}.\hskip 1em plus 0.5em minus 0.4em\relax {IEEE}, 2024. [Online]. Available: \url{https://doi.ieeecomputersociety.org/10.1109/SP54263.2024.00047}
\BIBentrySTDinterwordspacing

\bibitem{DBLP:conf/eurosp/WinderixMP21}
H.~Winderix, J.~T. M{\"{u}}hlberg, and F.~Piessens, ``{Compiler-Assisted Hardening of Embedded Software Against Interrupt Latency Side-Channel Attacks},'' in \emph{2021 {IEEE} European Symposium on Security and Privacy ({EuroS{\&}P})}.\hskip 1em plus 0.5em minus 0.4em\relax {IEEE}, 2021, pp. 667--682.

\bibitem{wolfssl21patches}
\BIBentryALTinterwordspacing
WolfSSL, ``{Ciphertext Side-Channel Patches}.'' [Online]. Available: \url{https://github.com/wolfSSL/wolfssl/pull/4666}
\BIBentrySTDinterwordspacing

\end{thebibliography}
}

\appendices

\appsection{\DmpsTitle{}}
\label[appsec]{app:dmp}

Apart from the previously discussed leakages originating from \ctscs{} and \ssss{}, \dmps{} (\DMPs{}) also fall into the category of \scname{} side-channels.
In this section, we introduce the \DMP{} leakage patterns and discuss to what extent interleaving can be used as a mitigation against the leakage.

\subsection{Leakage from \DmpsTitle{}}
\Dmps{} (\DMPs{}) are a class of prefetchers that prefetch data based on the program's data memory~\cite{DBLP:conf/isca/VicarteSNT0KF21, intelDdp}.
This is in contrast to ``classical'' prefetchers that are based on regularly occurring memory access patterns.
In order to also capture and speed up memory access patterns that involve layers of indirection or pointer chasing, the memory content is taken into account for prefetching program data.
If a memory read fulfills the \DMP{} activation criteria, the \DMP{} starts dereferencing data found at the data load address.
The \DMP{} should be disabled when enabling data (operand) independent timing (DOIT) mode~\cite{intelDdp,DBLP:conf/uss/ChenWSFKPG24}.
However, there is no kernel support yet and if the DOIT state can be toggled by the hypervisor, further software protection is needed.

\begin{figure}
    \centering
    \includegraphics[width=\linewidth]{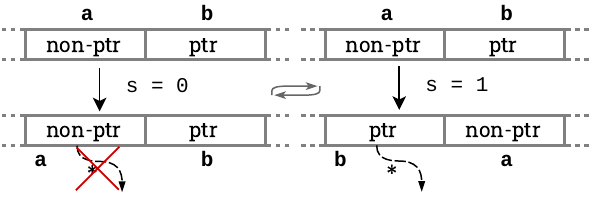}
    \caption{\Dmp{} (\DMP{}) leakage in \cswap{}. For a chosen-input-attack (described in \gofetch{}~\cite{DBLP:conf/uss/ChenWSFKPG24}) the attacker fills one of the arrays, in this case~\texttt{b}, with pointer-like data. They then set up the \DMP{} to dereference the data at the positions of the other array, and launch a cache attack to monitor whether the data at the first of the two addresses has been dereferenced by the \DMP{} as a result of a memory read (depicted by the dashed arrows). Thereby, the attacker can infer that the secret bit \texttt{s} was set if and only if the dereferencing succeeded.}
    \label{fig:cswap-dmp}
\end{figure}

Cases of data-at-rest leakage through a pointer-chasing \DMP{} have been shown for Apple's M series~\cite{DBLP:conf/uss/ChenWSFKPG24}, and for Intel's 13th Gen Raptor Lake microarchitecture, the existence of a \DMP{} with different activation criteria was demonstrated~\cite{wang2025peekPreprint}.
While concrete implementations of \DMPs{} differ, the exemplary pointer-chasing \DMP{} versions are based on scanning continuous 8-byte chunks and trying to dereference data found in those chunks.
\Cref{fig:cswap-dmp} shows the leakage from \cswap{} through the \DMP{} analyzed in~\cite{DBLP:conf/uss/ChenWSFKPG24}.

In the \DMP{} leakage scenario, not only a distinction between secret and public data of the victim is important, but also the impact of attacker-controllable data on other computations has to be considered.
The prefetcher that can be manipulated with the help of attacker-controllable data influences the \uarch{} state, so that secret data or information about victim processes can be leaked through side-channel attacks, even without any direct interaction with the adversarial code running~\cite{DBLP:conf/sp/VicarteFPG0FK22, DBLP:conf/uss/ChenWSFKPG24}.

\subsection{Preventing Leakages from \DmpsTitle{} via Interleaving}
\label{sec:prev-dmp} 

With suitable instrumentation in place, all \scname{} leakages that arise from \emph{writing} repetitive data to memory (see~\Cref{sec:background}) can be avoided.
However, there also are tight relations between \ctscs{} and \emph{read}-based leakages like \dmping{} (\DMP{}).
In this section, we analyze interconnections between \ctsc{} and \DMP{} attacks, and outline to what extent interleaving may be adapted for countering \DMP{} leakage.
Our threat model assumption for these considerations is the same as in~\Cref{sec:prev-sss}.

\paragraph{\CtscTitle{} Mitigation vs. \DMP{}}
The key idea to preventing \DMP{} leakage is breaking the observable relation between data and the corresponding \DMP{} activation. 
\Ctsc{} mitigations that add masking to secret plaintexts before writing to memory are thus effective, if the data is randomized.
However, for providing a sufficient security level, existing masking approaches like \cf{} incur high overheads due to the expensive random number generation~\cite{DBLP:conf/uss/WichelmannPW023}.
The 16-byte block interleaving in \tool{} that promises lower performance overhead cannot be used directly:
Pointers are typically aligned to 8-byte boundaries, so they are unaffected by the additional counter in every second 8-byte chunk.
If the \DMP{} still triggers for the new memory layout, it can directly dereference these pointers, ignoring the counters altogether.

\paragraph{Extending Interleaving to \DMP{}}
If interleaving is to be used for \DMP{} prevention, a smaller granularity is needed to break up \DMP{} activation patterns, effectively reducing the block size from 16 bytes to 8 bytes.
Instead of dividing data into 8-byte chunks and interleaving them with 8-byte counters, the data and counter chunk sizes need to be shrunk to implementation-dependent smaller chunks, such as 4 bytes each.
By that, the supported data size is reduced to 4 bytes, which causes significant problems:
8-byte integers, which are mandated by the architecture and widely used in cryptographic libraries, need to be split into two 4-byte parts.
Such split integers can no longer be read through a standard memory load, but need to be emulated.
Additionally, the changes affect a primitive type that is assumed to exist throughout the compiler framework, so tweaks need to be applied at all layers.
Due to the significant engineering effort, we leave the implementation of the exploration of this interleaving adaption for future work.

\begin{figure}
    \centering
    \includegraphics[width=\linewidth]{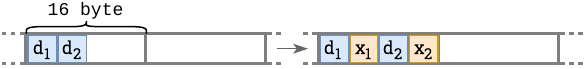}
    \caption{Interleaving data and values for freshness and \DMP{} protection in memory. In order to ensure protection for each 8-byte block in memory, the blocks are split into two chunks of 4 byte each, and contain either data (\texttt{d_i}) or freshness and \DMP{} protection values (marked as~\texttt{x_i}).}
    \label{fig:zebra-dmp}
\end{figure}

In addition to supporting another block size, the former freshness value has to be replaced by a combination of counter and \DMP{} invalidation.
For pure \DMP{} mitigation, the counters do not need to be incremented and can be set to a fixed value that prevents \DMP{} activation.
If the mitigation is meant also protect against other classes of \scname{} side-channel leakage, the counters can be incremented as before while ensuring that no values occur that again trigger the \DMP{}.
An illustration of the new memory layout can be found in~\Cref{fig:zebra-dmp}. 
An implementation needs to be careful with endianness, as to ensure that the counter ends up transforming the necessary bytes of a potential pointer. 

In this mitigation description, we assume that only canonical user-level pointer candidates, i.e., virtual addresses that may trigger prefetching, leak secret information via getting dereferenced. 
This is in line with current findings~\cite{wang2025peekPreprint}, that exclude leakage of non-canonical or kernel space addresses. 
With interleaving-based methods, mitigations may targets the root cause of the leakage, i.e., the \DMP{} activation on a data chunk.
Thereby, the aim is to also prevent subsequent exploitable leakages stemming from side-channels based on address translation or TLB lookups~\cite{wang2025peekPreprint}.
In addition, we assume the worst case leakage scenario, i.e., that the microarchitecture is already in a state that would immediately leak information when a pointer candidate is selected for prefetching.
Note that we base the explanation of extensibility on the attack vectors presented in~\cite{wang2025peekPreprint}, combined with theoretical reasoning about address translation properties.
Concrete implementations of \uarchopts{} may vary regarding specific processor or architectural features.
We therefore conclude that for an efficient and secure mitigation against \DMP{} leakage, the usage of verifiable DOIT/DIT modes is more effective if those are available.     

\end{document}